\documentclass[pra,twocolumn]{revtex4-2}
\usepackage{amsmath,amssymb,mathrsfs}
\usepackage{psfrag}
\usepackage{graphicx}
\usepackage{graphics}
\usepackage{epsfig}
\usepackage{bm}
\usepackage{color}
\usepackage{verbatim,color}
\usepackage{physics}
\usepackage[normalem]{ulem}
\usepackage[acronym]{glossaries}
\usepackage[hidelinks]{hyperref}
\usepackage{braket}

\newcommand{\beq}{\begin{equation}}
\newcommand{\eeq}{\end{equation}}
\newcommand{\beqa}{\begin{eqnarray}}
\newcommand{\eeqa}{\end{eqnarray}}
\newcommand{\ba}{\begin{aligned}[b]}
\newcommand{\ea}{\end{aligned}}

\usepackage[dvipsnames]{xcolor}

\begin{document}
\title{Repulsive fermions and shell effects on the surface of a sphere}
\author{L. Frigato$^{1,2}$, A. Bardin$^{1,2}$,  and L. Salasnich$^{1,2,3}$}
\affiliation{$^{1}$Dipartimento di Fisica e Astronomia ``Galileo Galilei", 
Universit\`a di Padova, Via Marzolo 8, 35131 Padova, Italy\\
$^{2}$Istituto Nazionale di Fisica Nucleare (INFN), Sezione di Padova, via Marzolo 8, 35131 Padova, Italy\\
$^{3}$Istituto Nazionale di Ottica (INO) del Consiglio Nazionale delle Ricerche (CNR), via Nello Carrara 1, 50019 Sesto Fiorentino, Italy}

\date{\today}
\begin{abstract}
 In recent years, ultracold atomic gases confined in curved geometries have attracted considerable theoretical interest. This is motivated by recent realizations of bubble traps in microgravity conditions, which open the possibility of investigating quantum many-body physics beyond the conventional flat-space paradigm. The theoretical interest up to now was mainly focused on Bose gases and their phenomenology, and has left the study of Fermi gases behind. In this paper, we investigate a two-component repulsive Fermi gas constrained to the surface of a sphere at finite temperature. We first analyze the non-interacting case, showing how the intrinsic geometrical features of the spherical surface give rise to a shell structure and modify the low-temperature thermodynamics compared to the flat two-dimensional gas. Repulsive interactions are then considered through an effective path-integral approach within a Hartree-Fock mean-field approximation, enabling us to derive the grand canonical potential and to regularize the associated Matsubara summation. We then investigate the stability of the spin-balanced state and obtain the finite-temperature Stoner criterion for fermions on a sphere, highlighting the interplay between the repulsive interactions and shell effects.

\end{abstract}
\maketitle
\section{Introduction}

Ultracold atomic gases have long attracted considerable interest, both theoretically and experimentally, as they provide an exceptional platform to explore many-body quantum physics \cite{Bloch2008, Bloch2017}. The ability to tune interatomic interactions via Feshbach resonances \cite{OHara, Feshbach1, Feshbach2, Feshbach3, Feshbach4} and the possibility to avoid solid-state system complications (such as phonons and impurities), along with the experimental developments in atomic confinement and cooling techniques, nowadays enable the study of a wide range of phenomena. This unique level of control has led in the past to groundbreaking experimental achievements with both bosonic and fermionic atomic gas mixtures (see, e.g. \cite{Anderson1995, Ketterle1995, DeMarco1999, Madison2000, Greiner2002, Regal2004, Zwierlein2004, Kinast2004, Zwierlein2005, Hadzibabic2006}) and establishes a central role for ultracold atoms in the development of future quantum computers and simulators \cite{Bloch2017, Cornish2024}.

In this context, recent experimental advancements now make it possible to trap ultracold atoms in very peculiar curved low-dimensional configurations, such as thin spherical or ellipsoidal shells \cite{Lundblad2019, Guo2022, Carollo2022, Jia2022}.  However, while the theoretical understanding of bosonic gases in such curved manifolds is relatively well advanced (see \cite{Tononi2023, Tononi2024} for comprehensive reviews), the investigation of fermionic gases remains less developed. Recent works in this sense have addressed non-interacting electrons \cite{Electronssphere}, the BCS-BEC crossover \cite{BCSBECsphere}, vortex structures \cite{vortexsphere} and phase separation on the surface of a sphere \cite{phasesepsphere}.

Inspired by the realization of atomic bubble traps in microgravity conditions at the Cold Atoms Laboratory onboard the International Space Station \cite{Carollo2022}, this paper studies a two-component gas of repulsive fermions confined to the surface of a sphere at finite temperature, making use of a suitable effective path integral formalism. A spherical surface, even if conceptually simple, exhibits non-trivial topological (compactness) and geometrical (presence of a constant curvature) features, which are reflected in the physical properties of the system, such as the need for periodic boundary conditions for the atomic motion \cite{Tononi2023}. Furthermore, contrary to what happens for bosons, which can condense \cite{BECsphere} and therefore macroscopically occupy the ground state of the system, for fermions this is not possible - unless there is an effective attractive interaction between them - due to the Pauli exclusion principle. This gives rise, on a sphere, to shell effects due to the organization of atoms into shells labeled by the angular momentum quantum number, which provides the most natural basis to describe the system \cite{BECsphere, BCSBECsphere, phasesepsphere}.

Moreover, when a repulsive interaction between the atoms is present, the system may undergo spontaneous polarization due to the competition between kinetic and interaction energy according to Stoner theory \cite{Stoner, Stonerth1, Stonerth2, Stonerth3, Stonerth4, Stonerth5}. Even though a direct experimental observation of this fact has long been the subject of debate \cite{Stonerexp1, Stonerexp2, Stonerexp3, Stonerexp6, Stonerexp4, Stonerexp5}, recent experiments suggest that indeed a gas of fermions may exhibit spontaneous polarization \cite{Valtolina2017} under specific initial conditions. In this paper, we found that the interplay between such polarization tendency and the distinctive geometric features of the spherical manifold gives rise to nontrivial shell effects.

The paper is organized as follows. In Section \ref{Non-interacting}, we start by considering the dispersion relation of a single non-interacting particle confined to the surface of a sphere. Then, we derive the main thermodynamic quantities of interest for an ideal Fermi gas on a sphere, comparing them to the standard known results of the two-dimensional flat case. Interactions between fermions are taken into account in Section \ref{Repulsive fermion gas}, adopting a path integral formalism. In particular, the grand canonical potential of the system is derived within a mean-field HF approximation by explicitly performing the Gaussian functional integration and regularizing the corresponding divergent Matsubara frequency summation. Finally, in Section \ref{Stoner instability}, the stability of the spin-balanced solution is investigated through bifurcation theory, finding a phase transition modulated by the intensity of the interaction strength.  The finite-temperature Stoner criterion for the stability of a fermionic system on the sphere is thus derived within this mean-field approximation and compared to the standard two-dimensional flat result. The conclusions, with some experimental proposals, are presented in Section \ref{Conclusions} and conclude our work.

\section{Non-interacting fermion gas}
\label{Non-interacting}
Let us consider a two-component Fermi gas confined to the surface of a sphere and in thermal equilibrium at temperature $T$. Thermal equilibrium is experimentally achieved before the atoms are loaded, through radio-frequency potentials, in the final bubble trap configuration. In this preparation stage, the atoms are cooled using established cooling techniques, which allow the mixture to thermalize at very low temperatures \cite{Carollo2022}. Assuming a weakly interacting gas where thermalization is achieved, we first treat the ideal non-interacting case.

The energy of a particle of
mass $m$ constrained to move on the surface of a sphere of radius $R$ is quantized according to \cite{Electronssphere, BECsphere}
\begin{equation}
    \mathcal{E}_l=\frac{\hbar^2}{2mR^2} l(l+1),
    \label{single part energy}
\end{equation}
where $\hbar$ is the reduced Planck constant and $l \in \mathbb{N}$ is
the angular momentum quantum number. Notice how the spacing between the energy levels is inversely proportional to $R^2$. Each energy level (or shell) labeled by $l$ has a degeneracy of $2l+1$ due to the magnetic quantum number, which describes the projection along the $z$-axis of the angular momentum $l$, $m_l \in [-l, l]$. 

 According to quantum statistical mechanics \cite{Huang1987}, the average total number of non-interacting fermions (let us assume $N_\uparrow=N_\downarrow$= N/2) on the surface of a sphere of radius $R$ is
\begin{equation}
    N= \sum_{\sigma=\left\{ \uparrow, \downarrow \right\}} \sum_{l=0}^{+\infty} \frac{2l+1}{e^{\beta(\mathcal{E}_l-\mu_\parallel)}+1},
    \label{Average number}
\end{equation}
where $\sigma=\{\uparrow,\downarrow \}$ is the spin of the fermion and $\mu_\parallel$ is the chemical potential ($\mu^\parallel_\uparrow=\mu_\downarrow^\parallel=\mu_\parallel$), while $\beta=1/k_BT$ where $T$ is the system temperature and $k_B$ the Boltzmann constant. Inside Eq. (\ref{Average number}) we recognize the familiar Fermi-Dirac distribution: this suggests that, due to the Pauli exclusion principle, the fermions arrange themselves inside the degenerate angular momentum shells of the non-interacting single particle spectrum Eq. (\ref{single part energy}) in such a way that each $l$-shell can contain at most $2(2l+1)$ fermions.

When $\beta \to \infty$, it is well known that the Fermi-Dirac distribution reduces to the Heaviside function $\Theta(\mu-\mathcal{E}_l)$. The sum over the angular quantum number $l$ is therefore truncated up to a Fermi angular momentum $l_F$, so that $l_F$ corresponds to the highest occupied shell. Explicitly, at zero temperature
\begin{equation}
    N=\sum_{\sigma=\left\{ \uparrow, \downarrow \right\}} \sum_{l=0}^{l_F} (2l+1) = 2(l_F+1)^2.
    \label{Number T=0}
\end{equation}
Therefore, there are special values of $N$ for which shells at $T=0$ are completely closed, $N=2,8,18, 32 ... $, which are usually called magic numbers \cite{Electronssphere}.
The Fermi energy of the system is defined as
\begin{equation}
\varepsilon_F=\frac{\hbar^2}{2mR^2}l_F(l_F+1),
\label{Fermi energy}
\end{equation}
 and corresponds to the energy of the topmost occupied (degenerate) energy level. Consequently, the Fermi angular momentum is determined by 
\begin{equation}
    l_F=\left\lfloor -\frac{1}{2}+\frac{1}{2}\sqrt{1+\frac{8mR^2}{\hbar^2} \varepsilon_F}\right\rfloor,
    \label{l_fermi intero}
\end{equation}
where $\lfloor x \rfloor$ is the integer part of $x$. We stress that the presence of the``floor'' function $\lfloor x \rfloor$ is crucial: since the Fermi angular momentum must be $l_F \in \mathbb{N}$, neglecting it would prevent us from obtaining the correct shell structure. Notice also how Eq. (\ref{l_fermi intero}) is equivalent to the condition imposed by the Heaviside function $\Theta(\mu-\mathcal{E}_l)$. These results are equivalent to those discussed for electrons in \cite{Electronssphere}.

\begin{figure}
    \label{fig:placeholder}
\includegraphics[width=1\linewidth]{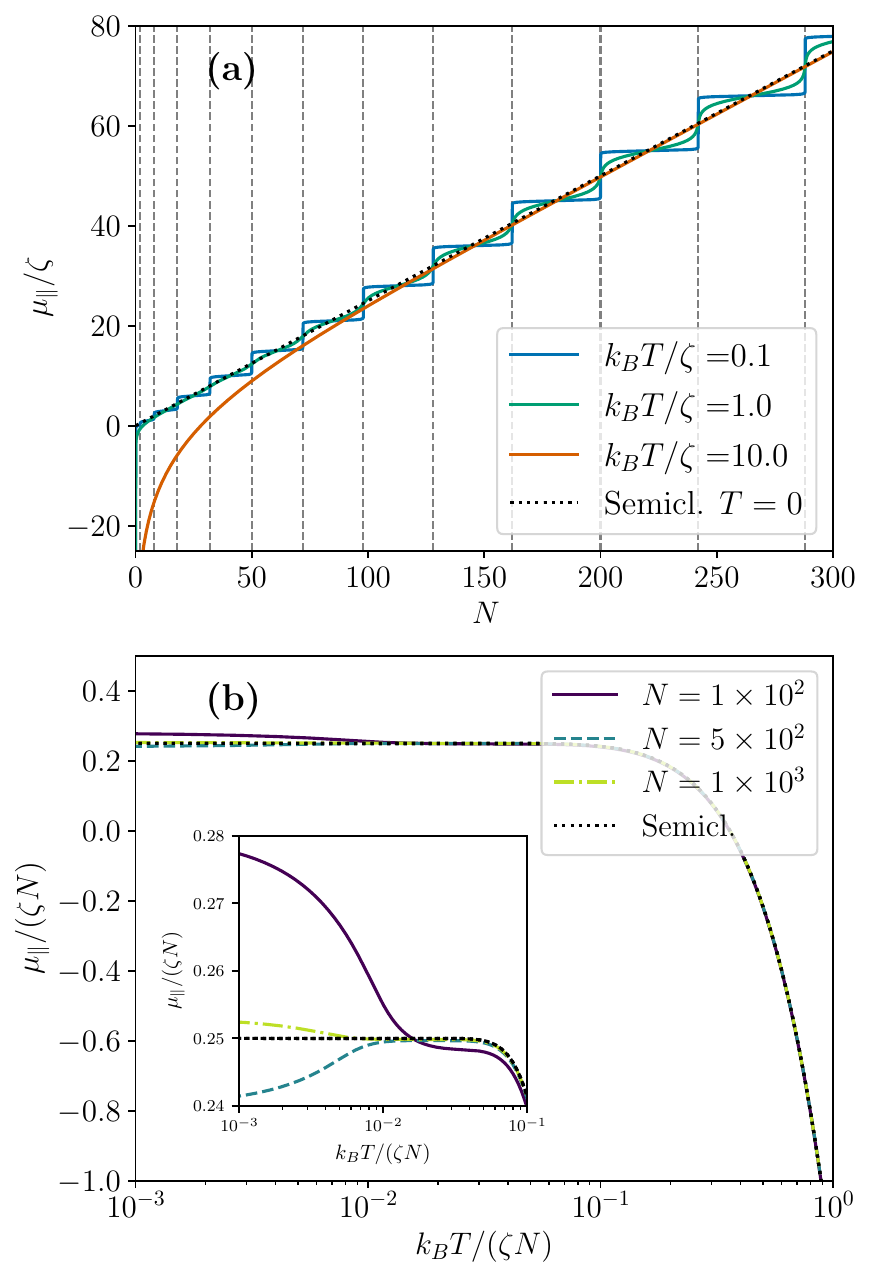}
\caption{Thermodynamics of the non-interacting Fermi gas. All the quantities are made dimensionless by introducing the sphere energy scale $\zeta=\hbar^2/(mR^2)$, as explained in the main text. (a) Dimensionless chemical potential $\mu_\parallel/\zeta$ as a function of the number of fermions $N$ for different temperatures, namely $k_BT/\zeta=10$ (orange solid line), $k_BT/\zeta=1.0$ (green solid line) and $k_BT/\zeta=0.1$ ( blue solid line). 
The gray vertical dashed lines mark the magic numbers, corresponding to completely filled shells. (b) Chemical potential per particle $\mu_\parallel(T)/(N \zeta)$ as a function of the dimensionless temperature scaled for the total number of particles, $k_BT/(\zeta N)$, for different system sizes: $N=10^2$ (purple solid line), $N=5 \times10^{2}$ (dashed light blue) and  $N=10^3$ (dash-dotted light green). The semiclassical limit (black dotted line) is also plotted for comparison.}
    \label{fig 1}
\end{figure}

In Fig. \ref{fig 1}-(a), we plot the dimensionless chemical potential as a function of the average number of fermions $N$ to illustrate the gas shell structure, which arises from the quantization of angular momentum, and its dependence on the temperature of the system, as governed by the Fermi-Dirac distribution. For sufficiently low temperatures, the average number exhibits a clear step-like behavior, revealing the underlying shell structure due to the angular momentum algebra, while for sufficiently high values of $T$ the step-like behavior is washed out. Crucially, for completely filled shells (i.e., at magic numbers values) in the $T \to 0$ limit, the chemical potential does not simply coincide with the Fermi energy $\varepsilon_F$. Due to the energy gap that separates the Fermi energy level from the lowest unoccupied one, the chemical potential can rigorously assume any value strictly within this gap. If the $l_F$-shell is only partially filled, the chemical potential instead coincides exactly with the Fermi energy, $\mu(T=0) = \varepsilon_F$.

In full generality, we can also determine the non-interacting grand canonical potential, which encodes all the equilibrium thermodynamics of the system \cite{Huang1987}. It reads
\begin{equation}
     \Omega_0= - \frac{1}{\beta} \sum_{\sigma=\{\uparrow, \downarrow \}} \sum_{l=0}^{+\infty} (2l+1) \ln \left[e^{-\beta(\mathcal{E}_l - \mu_\parallel)}+1 \right]
     \label{Grand potential free}.
 \end{equation}
For example, Eq. (\ref{Average number}) could have been derived from Eq. (\ref{Grand potential free}) using the standard thermodynamics relation $N=-\partial \Omega_0/\partial \mu_\parallel$, the gas entropy can be obtained as $S=-\partial \Omega_0/\partial T$ and its pressure is given by $P=-\partial \Omega_0/\partial V$. 

Before moving on, it is important to make a remark. The system under study is closed and finite, with a conserved particle number (neglecting small losses due to three-body processes or trap imperfections). While this makes the grand canonical ensemble seemingly inappropriate, its use is justified because canonical approaches, though formally correct, are impractical due to the known mathematical difficulties when calculating the partition function. The grand canonical ensemble, instead, enables analytic progress and, for the temperature range we are interested in, provides a good approximation, as the relative error introduced by operating in it scales as $\propto 1/N^{3/4}$ (see Appendix \ref{Appendix D}).

Due to the presence of summations, Eqs. (\ref{Average number}) and (\ref{Grand potential free}) are rather unintelligible because no exact analytical solution can be obtained. To circumvent this problem, we can operate within the semiclassical approximation: the discrete sum over angular momenta is replaced by an integral, $ \sum_{l=0}^{+\infty} \longrightarrow \int_0^{+\infty} dl $, allowing us to obtain analytical results. This approximation becomes exact in the $R \to +\infty$ limit, where the single-particle spectrum Eq. (\ref{single part energy}) becomes continuous. This corresponds to the thermodynamic limit of the system, where also $N \to +\infty$ such that $n = N/V=const$, so that only intensive quantities remain well defined. Notice that the hyper-volume $V$ here actually corresponds to the total area of the spherical surface, $V=4\pi R^2$. Proceeding in this way, Eq. (\ref{Average number}) becomes
\begin{equation}
    n=\frac{N}{V}=\frac{m}{\pi\beta\hbar^2 }\ln(1+e^{\beta \mu_\parallel}),
    \label{density semiclassical sphere}
\end{equation}
which can be explicitly inverted (as opposed to Eq. (\ref{Average number})), yielding the chemical potential as a function of the total number density of the system
\begin{equation}
    \mu_\parallel= \frac{1}{\beta} \ln\left[ e^{\left( \frac{\pi \beta \hbar^2 }{m} n \right)} - 1 \right]
    \label{mu as funct of n semiclass};
\end{equation}
while from Eq. (\ref{Grand potential free}), the semiclassical non-interacting grand potential density $\omega_0$ reads 
 \begin{equation}
   \omega_0=\frac{\Omega_0}{V} =  \frac{m}{\beta^2\hbar^2\pi} \operatorname{Li}_2(-e^{\beta \mu_\parallel}),
    \label{Gran Potential semiclass 1}
\end{equation}
where $\operatorname{Li}_2(x)$ is the polylogarithmic function and $V=4\pi R^2$ is the surface of the sphere.

Unlike bosons, for which it is natural to separate the condensate - the lowest energy level - from the excited states, no analogous decomposition is meaningful for fermions. As a result, taking the limit $R \to +\infty$ inevitably reduces the problem to that of a flat infinite two-dimensional system. For fermions, the semiclassical approximation cannot capture curvature effects (such as the finite size of the sphere), as shown instead in Ref. \cite{BECsphere} for bosons, because any explicit dependence on $R$ is lost. An accurate description therefore necessarily requires retaining the discrete nature of the single-particle spectrum on the sphere: the information on the finite size of the system will be encoded within the summations over the discrete angular momenta $l$. Unfortunately, this comes with a price, as the presence of the sums prevents us from obtaining analytical results.

In Fig. \ref{fig 1}-(b) we compare the flat two-dimensional equation of state $\mu_\parallel(n, T)$ - provided by the semiclassical approximation - with the exact result on the sphere, $\mu_\parallel(n, T, R)$. The latter is obtained by numerically inverting Eq. (\ref{Average number}), which is in turn also evaluated numerically (see Appendix \ref{Appendix B} for more details), fixing the average number of fermions $N$ on the sphere. In particular, we plot the chemical potential per particle, $\mu_\parallel/N$, which is made dimensionless by introducing the sphere energy scale $\zeta=\hbar^2/(mR^2)$. Accordingly, all the other quantities are also made dimensionless, so that Eq. (\ref{mu as funct of n semiclass}) assumes the dimensionless form
\begin{equation}
    \frac{\mu_\parallel}{\zeta N}
=
\frac{k_B T}{\zeta N}
\ln\!\left[
e^{\left(\frac{\zeta N}{4 k_B T}\right)} - 1
\right].
\end{equation}
In this way, since $nR^2=N/(4\pi)$ (dimensionless), and given that the zero-temperature limit of Eq. (\ref{mu as funct of n semiclass}) yields $\varepsilon_F= \hbar^2 \pi n/m$, then semiclassically $\varepsilon_F/(N \zeta) = 0.25 $, regardless of the number of fermions $N$. At fixed $N$, we observe deviations between the semiclassical solutions ( black dotted line) and exact (colored solid and dashed lines) at low temperatures, while convergence to the semiclassical solution is recovered at high temperatures. In the low temperature regime, the semiclassical approximation fails to accurately describe the exact behavior since the discreteness of the spectrum in this case cannot be ignored. Such discretization of the energy levels is an intrinsic property of the spherical surface (due to the finite size of the system) and gives rise to non-trivial shell effects. Depending on whether the Fermi level is more or less than half-filled, the chemical potential either rises or drops sharply as soon as $T \ne 0$ to conserve the total particle number as $N$ is kept fixed. At higher temperatures, once several excited states become thermally accessible, all curves gradually approach the semiclassical behavior. Moreover, as the number of atoms increases, the discrepancy between the exact discrete summation and the semiclassical continuous approximation decreases throughout the temperature range.

\section{Repulsive fermion gas}
\label{Repulsive fermion gas}

Let us now consider a gas of repulsive interacting fermions on the surface of a sphere in thermal equilibrium at a temperature T. Within a path integral formalism (see Appendix \ref{Appendix A}), the grand canonical partition function of the system is given by
\begin{equation}
    Z=\int \mathcal{D}[\bar{\psi}_\sigma, \psi_\sigma] \ e^{-S_E[\bar{\psi}_\sigma, \psi_\sigma]/\hbar},
    \label{path integral}
\end{equation}
where
\begin{equation}
    S_E[\bar{\psi}_\sigma, \psi_\sigma]=\int_0^{\hbar\beta} d\tau \int_0^{2\pi} d\varphi \int_0^\pi \sin\theta d\theta \ \mathcal{L}_E[\bar{\psi}_\sigma, \psi_\sigma]
    \label{Euclidean action}
\end{equation}
is the Euclidean action and    
\begin{equation}
\begin{aligned}
  \mathcal{L}_{E}[ \bar{\psi}_\sigma, {\psi}_\sigma ]=&\sum_{\sigma=\{\uparrow, \downarrow \}} \bar{\psi}_\sigma \left( \hbar\frac{\partial}{\partial \tau}+\frac{\hat{L}^2}{2mR^2}-\mu^{\parallel}_\sigma \right){\psi}_\sigma\\ &-g_{\parallel} \bar{\psi}_\uparrow  \bar{\psi}_\downarrow {\psi}_\downarrow {\psi}_\uparrow 
  \label{full eucl lagr}
\end{aligned}    
\end{equation}
is the Euclidean Lagrangian density per solid angle, namely the Lagrangian density in imaginary time $\tau=it$. Notice that the kinetic energy of the atoms is proportional to $\hat{L}^2$, the square of the angular momentum operator, which can be expressed in spherical coordinates 
\begin{equation}
    \hat{L}^2=-\hbar^2 \left[ \frac{1}{ \sin\theta}\frac{\partial}{\partial \theta}\left( \sin\theta \frac{\partial}{\partial \theta }\right)+\frac{1}{ \sin^2\theta} \frac{\partial^2}{\partial \varphi^2 } \right] 
    \label{L^2 operator}
\end{equation}
as the particles are constrained on the surface of a sphere. The Grassmann fields $\bar{\psi}_\sigma$, $\psi_\sigma$ appearing in Eq. (\ref{full eucl lagr}) are dimensionless since the integration in Eq. (\ref{Euclidean action}) is over the solid angle, and describe fermions moving on the surface of a sphere interacting through a repulsive contact potential of strength $g_{\parallel} > 0$. Now we also allow for a population imbalance between the two spin components, $N_\uparrow \ne N_\downarrow$, so that, in general, $\mu^{\parallel}_\uparrow \ne \mu^{\parallel}_\downarrow$.

To treat the interacting quartic term in Eq. (\ref{full eucl lagr}), we implement a mean-field Hartree-Fock (HF) approximation. In this scheme, the fields are decoupled as
$\bar{\psi}_\uparrow  {\psi}_\uparrow  = \braket{ \bar{\psi}_\uparrow  {\psi}_\uparrow } + \delta( \bar{\psi}_\uparrow  {\psi}_\uparrow ) $, $\bar{\psi}_\downarrow  {\psi}_\downarrow  = \braket{ \bar{\psi}_\downarrow  {\psi}_\downarrow} + \delta( \bar{\psi}_\downarrow  {\psi}_\downarrow )$, and we define $\tilde{n}_\uparrow=\braket{ \bar{\psi}_\uparrow  {\psi}_\uparrow }$, $\tilde{n}_\downarrow=\braket{ \bar{\psi}_\downarrow  {\psi}_\downarrow }$ as the number density per solid angle of the spin up and down fermions, respectively. Furthermore, we neglect the quantum fluctuations $\delta( \bar{\psi}_\uparrow  {\psi}_\uparrow)$, $\delta( \bar{\psi}_\downarrow  {\psi}_\downarrow) $, so that the resulting HF Euclidean Lagrangian density is quadratic in the fields
\begin{equation}
\begin{aligned}
  \mathcal{L}_{HF}[ \bar{\psi}_\sigma, {\psi}_\sigma ]=&\sum_{\sigma=\{\uparrow, \downarrow \}} \bar{\psi}_\sigma \Big( \hbar\frac{\partial}{\partial \tau}+\frac{\hat{L}^2}{2mR^2}-\mu^{\parallel}_\sigma\\ &+g_{\parallel}\tilde{n}_{-\sigma} \Big){\psi}_\sigma-g_{\parallel} \tilde{n}_\uparrow \tilde{n}_\downarrow  
\end{aligned}    
\end{equation}
and we can proceed with its functional integration. Notice that we have explicitly introduced the number densities $\tilde{n}_\uparrow$, $\tilde{n}_\downarrow$ inside the Lagrangian, which already contained the chemical potentials $\mu^{\parallel}_\uparrow$,  $\mu^{\parallel}_\downarrow$. Thus, at the end of the calculations, the grand canonical potential must be minimized with respect to these densities according to the variational principle, so that $\tilde{n}_\uparrow$ and $\tilde{n}_\downarrow$ can be determined self-consistently. 

The functional integration of the HF Lagrangian can be performed by expanding the field
\begin{equation}
\psi_\sigma(\theta,\varphi,\tau) = \sum_{l=0}^\infty \sum_{m_l=-l}^l \sum_{s=-\infty}^{+\infty} c_{l, m_l, \sigma}^{(s)} e^{-i\omega_s \tau} Y_{l}^{m_l}(\theta,\varphi),   
\end{equation}
and similarly for the Grassmann conjugated $\bar{\psi}_\sigma(\theta,\varphi,\tau)$,
where $Y_{l}^{m_l}(\theta,\varphi)$ are the spherical harmonics (they provide an orthonormal basis set for the single particle eigenvalue problem) and $\omega_s={(2s+1) \pi}/({\hbar \beta})$ the fermionic Matsubara frequencies, $s \in \mathbb{Z}$. Exploiting the
orthonormality properties of $Y_{l}^{m_l}(\theta,\varphi)$ and of the complex exponential, after integration, the Euclidean action reduces to
\begin{equation}
\begin{split}
    S_{HF}[\bar{\psi}_\sigma, \psi_\sigma]=\hbar \beta \sum_{\sigma=\{\uparrow,\downarrow\}} \sum_{l=0}^{+\infty} \sum_{m_l=-l}^l \sum_{s=-\infty}^{\infty} \bar{c}^{(s)}_{l, m_l, \sigma} \\ \times \left[ -i \hbar\omega_s  + \mathcal{E}_l - \mu^{\parallel}_\sigma +g_{\parallel} \tilde{n}_{-\sigma} \right]{c}^{(s)}_{l, m_l, \sigma}-4\pi\hbar\beta \ \tilde{n}_\uparrow \tilde{n}_\downarrow.
\end{split}    
\end{equation}
It is now convenient to define the quantity 
\begin{equation}
\lambda_{\sigma, l}^{(s)}=\beta(-i\hbar \omega_s + \varepsilon_{l, \sigma})    
\end{equation}
where 
\begin{equation}
\varepsilon_{l, \sigma}(\tilde{n}_{-\sigma})= \mathcal{E}_l-\mu^{\parallel}_\sigma+g_{\parallel}\tilde{n}_{-\sigma}    
\label{Hartree Fock spectrum}
\end{equation}
is the single-particle HF energy of the fermions. The partition function is thus given by performing the Gaussian Grassmann-Berezin integrals
\begin{align}
    Z_{HF}&=\int \prod _{\sigma, l, m_l, s} d\bar{c}^{(s)}_{\sigma, l, m_l} \ d{c}^{(s)}_{\sigma, l, m_l} \ e^{-{S_{HF}}/{\hbar}}\\ &= e^{4\pi\beta \ \tilde{n}_\uparrow \tilde{n}_\downarrow} \prod_{\sigma= \{\uparrow,\downarrow\}} \prod_{l=0}^{+\infty} \prod_{m_l=-l}^{l} \prod_{s=-\infty}^{+\infty} \lambda_{\sigma, l}^{(s)}.
    \label{HF partition function}
\end{align}
Taking the logarithm of Eq. (\ref{HF partition function}), we find the HF grand canonical potential
\begin{align}
     \Omega_{HF}&=-\frac{1}{\beta} \ln{Z_{HF}}\\ &= -\frac{1}{\beta}\sum_{\sigma=\{\uparrow,\downarrow\}} \sum_{l=0}^{+\infty} \sum_{m_l=-l}^l \sum_{s=-\infty}^{\infty} \ln \lambda_{\sigma, l}^{(s)} -4\pi g_{\parallel} \tilde{n}_{\uparrow}\tilde{n}_{\downarrow}.
\end{align}
 The sum over the Matsubara frequency is divergent, but this divergence is artificial and can be resolved by taking into account in the summation a convergence factor $e^{i\omega_s 0^+}$, which arises from the implicit time ordering in the path integral (see Appendix \ref{Appendix C}). Finally, we obtain the grand canonical potential within the HF approximation for a gas of repulsive fermions on the sphere
\begin{equation}
    \Omega_{HF}=-4\pi g_{\parallel} \ \tilde{n}_\uparrow \tilde{n}_\downarrow  -\frac{1}{\beta}\sum_{\sigma=\{\uparrow,\downarrow\}} \sum_{l=0}^{+\infty} (2l+1)  \ln \left( 1+e^{-\beta \varepsilon_{ \sigma,l}} \right)
    \label{HF grand potential}
\end{equation}
where we performed the sum over the magnetic quantum number, $\sum_{m_l=-l}^l=(2l+1)$. 
Notice how this expression reduces to Eq. (\ref{Grand potential free}) if $g_\parallel =0$. If $g_\parallel \ne0$, the single-particle energy Eq. (\ref{Hartree Fock spectrum}) presents an additional shift term, which corresponds to the HF mean-field contribution.

Minimizing the grand canonical potential with respect to the variational number densities
\begin{equation}
\frac{\partial \Omega_{HF}}{\partial \tilde{n}_{\sigma}}=0   
\label{Saddle point condition}
\end{equation}
corresponds to imposing the saddle point condition we implicitly employed to solve the path integral (\ref{path integral}) within the HF mean field approach. The direct evaluation of the saddle point condition yields the following set of coupled equations
 \begin{equation}
\left\{
\begin{aligned}
{n}_\uparrow &= \frac{1}{4\pi R^2} \sum_{l=0}^{+\infty} \frac{(2l+1)}{1 + e^{\beta(\mathcal{E}_l-\mu^{\parallel}_\uparrow+g_{2D}{n}_\downarrow)  }} \\
{n}_\downarrow &= \frac{1}{4\pi R^2} \sum_{l=0}^{+\infty} \frac{(2l+1)}{1 + e^{\beta(\mathcal{E}_l-\mu^{\parallel}_\downarrow+g_{2D}{n}_\uparrow) }}
\end{aligned}
\right.
\label{HF equilibrium densities}
\end{equation}
which describes the average equilibrium densities of the spin populations. These two equations are exactly those obtained by the standard relation ${n}_\sigma=-(1/V)(\partial \Omega_{HF}/\partial \mu^\parallel_\sigma) $, so this fully justifies identifying $\braket{\bar{\psi}_\sigma\psi_\sigma}$ with the average densities. The two equations are coupled by the interaction strength $g_{2D}$ and must be solved self-consistently, since $n_\sigma$ appears both on the left-hand side and in the exponent on the right-hand side of the equations. Notice that we have dropped the tilde, since now we express the density on the surface of a radius sphere $R$ ($n_\sigma=\tilde{n}_\sigma/ R^2$), rather than per solid angle.  After rescaling all quantities according to the above change of variables, the two-dimensional interaction strength
\begin{equation}
    g_{2D}=g_{\parallel}R^2,
    \label{g_2D definition}
\end{equation}
naturally appears in Eq. (\ref{HF equilibrium densities}) (see Appendix \ref{Appendix A}).

At fixed temperature $T$, the densities $n_{\sigma=\{\uparrow,\downarrow\}}$  that satisfy the saddle point condition (\ref{Saddle point condition}) are fully determined by the chemical potentials $\mu^{\parallel}_{\sigma=\{\uparrow,\downarrow\}}$ and the interaction strength $g_{2D}$. If $\mu^{\parallel}_\uparrow=\mu^{\parallel}_\downarrow=\mu_{\parallel}$, this set of coupled equations always admits the symmetric solution $n_\uparrow=n_\downarrow=n/2$. Observe also that the system is invariant under the exchange $n_\uparrow \leftrightarrow n_\downarrow$, $\mu^{\parallel}_\uparrow \leftrightarrow \mu^{\parallel}_\downarrow$.

For completeness, let us analyze the $\beta \to +\infty$ limit of Eq. (\ref{HF equilibrium densities}). This time, we find that the Fermi angular momentum for the $\sigma$-species is given by
\begin{equation}
    l_F^{\sigma}=\left\lfloor -\frac{1}{2}+\frac{1}{2}\sqrt{1+\frac{8mR^2}{\hbar^2} (\varepsilon_F^\sigma-g_{2D}n_{-\sigma})}\right\rfloor
    \label{l_fermi intero interaction 2}.
\end{equation}
while the Fermi energy $\varepsilon^\sigma_F$ generalizes to 
\begin{equation}
    \varepsilon^\sigma_F=\frac{\hbar^2}{2mR^2}l^\sigma_F(l^\sigma_F+1)+g_{2D}n_{-\sigma}
    \label{Fermi energy interacting unb}
\end{equation}
where now $l^\sigma_F$ is provided by Eq. (\ref{l_fermi intero interaction 2}) and the interactions are taken into account by the presence of the mean field interaction term, $g_{2D}n_{-\sigma}$.

Furthermore, proceeding as outlined in the non-interacting case, the semiclassical chemical potential can be obtained from Eq. (\ref{HF equilibrium densities}) 
\begin{equation}
    \mu_\sigma^{\parallel} = \frac{1}{\beta} \ln\left( e^{\frac{2\pi\beta \hbar^2 }{m} n_\sigma} - 1 \right)+g_{2D} n_{-\sigma}.
    \label{HF chemical pot}
\end{equation}

\section{Stoner instability on the surface of a sphere}
\label{Stoner instability}

We are now interested in studying the stability of the gas in the spin-balanced configuration at finite temperature, i.e., when $n_\uparrow=n_\downarrow=n/2$ and $\mu_\uparrow^\parallel=\mu_\downarrow^\parallel=\mu_\parallel$, using bifurcation theory \cite{Stonerth1, SalasnichStoner}. For this purpose, it is convenient to define the vector $\vec{n}=({n}_\uparrow, {n}_\downarrow)$, such that Eq. (\ref{HF equilibrium densities}) can be recast in a more compact form \cite{Stonerth1, SalasnichStoner} as
\begin{equation}
    \vec{F}(\vec{n})=\vec{0}
\end{equation} 
where
\begin{align}
    \nonumber \vec{F}(\vec{n})=\Big(&{n}_\uparrow - \frac{1}{4\pi R^2} \sum_{l=0}^{+\infty} \frac{(2l+1)}{1 + e^{\beta(\mathcal{E}_l-\mu^\parallel_\uparrow+g_{2D}{n}_\downarrow)  }} ,\\ &{n}_\downarrow - \frac{1}{4\pi R^2} \sum_{l=0}^{+\infty} \frac{(2l+1)}{1 + e^{\beta(\mathcal{E}_l-\mu^\parallel_\downarrow+g_{2D}{n}_\uparrow) }} \Big) 
    \label{JAcobian formal}.
\end{align}
To study the stability of the equilibrium solution, we have to calculate the Jacobian of $\vec{F}(\vec{n})$. Given the definition of $\vec{F}$, note that this is actually the Hessian of the grand canonical potential $\Omega_{HF}$ with respect to the average number densities $n_\sigma$.  It is important to stress that while the system is described within the grand canonical ensemble, $\Omega_{HF}$ explicitly depends on the densities $n_\uparrow$ and $n_\downarrow$, which act as variational order parameters introduced by the mean-field decoupling and not, strictly speaking, as conjugate thermodynamic variables to the chemical potential. The stability against infinitesimal uniform spin fluctuations requires this effective potential to be minimized with respect to these variational parameters, and the value for which this condition is met corresponds to the equilibrium homogeneous densities on the sphere. Consequently, the stability condition is dictated by the requirement that the Hessian matrix of $\Omega_{HF}$ with respect to the densities $n_\sigma$ remain positive definite \cite{Stonerth1, SalasnichStoner}. Thus, the finite temperature Stoner instability criterion is given by the following condition
\begin{equation}
   \det\left( \frac{\partial \vec{F}(\vec{n})}{\partial\vec{n}}\right)_{|{n}_\uparrow={n}_\downarrow=\frac{n}{2}} \le 0.
   \label{equation jacobian}
\end{equation}
When the inequality is satisfied, the symmetric equilibrium solution $n_\uparrow = n_\downarrow=n/2$ loses stability. Due to the symmetry of Eq. (\ref{HF equilibrium densities}), this instability leads to a pitchfork bifurcation: at the critical point (namely, when the equality holds), the symmetric solution splits into two new stable equilibrium branches, where $n_\uparrow \ne n_\downarrow$, symmetric under exchange $n_\uparrow \leftrightarrow n_\downarrow$, $\mu^{\parallel}_\uparrow \leftrightarrow \mu^{\parallel}_\downarrow$. In such a case, the system  would find it energetically convenient to lower the population of one of the two species to minimize the interaction energy $g_{2D}n_{-\sigma}$. Explicitly, Eq. (\ref{equation jacobian}) reads
\begin{equation}
     \frac{g_{2D}\beta}{16 \pi R^2} \sum_{l=0}^{\infty} (2l+1)\  \operatorname{sech}^2\left( \beta \frac{\varepsilon_l}{2} \right)   \ge 1
    \label{STONER CRITERION}.
\end{equation}
where 
\begin{equation}
\varepsilon_{l}(n)= \mathcal{E}_l-\mu_{\parallel}+g_{2D}\frac{n}{2}  . 
\label{Hartree Fock spectrum symmetric}
\end{equation}
This criterion suggests the possibility of realizing an interaction-driven transition from a spin-balanced to a spin-polarized Fermi gas on a spherical surface. In the absence of interactions ($g_{2D}=0$), no mechanism exists to destabilize the balanced state, as our model does not include a spin-flip mechanism and the symmetric solution is always stable with respect to fluctuations in the population of the species. However, for $g_{2D}\neq 0$, the stability of the unpolarized configuration is governed by Eq. (\ref{STONER CRITERION}). 

It is now worth commenting on the physical nature of the instability predicted by Eq. (\ref{STONER CRITERION}). In our Hartree-Fock framework, we restricted our analysis to spatially uniform density profiles. Within this uniform assumption and operating in the grand canonical ensemble, loss of stability corresponds to macroscopic $\mathcal{O}(N)$ fluctuations of the global spin populations, as the system attempts to lower its interaction energy by fully polarizing, exchanging particles with the reservoir. However, in realistic experimental setups with ultracold atomic gases, as we have already discussed, there is no particle reservoir, and the number of particles is conserved. Therefore, in an experimental scenario, the system avoids the instability of the homogeneous balanced state not by undergoing global particle number fluctuations but rather by breaking the spherical spatial symmetry, forming spin domains on the sphere. This is what was found and discussed in Ref. \cite{phasesepsphere} at $T=0$. We shall return to this point later.

The critical value of the interaction strength beyond which the unpolarized gas loses stability is obtained when the equality in Eq. (\ref{STONER CRITERION}) holds, namely when
\begin{equation}
    g_{2D, c} = \frac{1}{\beta}\frac{16\pi R^2}{ \sum_{l=0}^{+\infty} (2l+1)\operatorname{sech}^2\left( \beta \frac{\varepsilon_l}{2} \right)}
    \label{Stoner criterion g critical}
\end{equation}
which is an equation that must be solved self-consistently simultaneously with Eq. (\ref{HF equilibrium densities}), setting $n_\uparrow=n/2=n_\downarrow$ and $\mu_\uparrow^\parallel=\mu_\parallel=\mu_\downarrow^\parallel$.

Performing the substitution $\sum_{l=0}^{+\infty} \longrightarrow \int_0^{+\infty}dl$, the semiclassical two-dimensional Stoner criterion is found
\begin{equation}
    \frac{g_{2D}m}{2\pi\hbar^2} \frac{1}{(1+e^{\beta(g_{2D}\frac{n}{2}-\mu_{\parallel})})} \ge 1.
    \label{Semiclass Stoner criterion}
\end{equation}
Combining Eqs. (\ref{Semiclass Stoner criterion}) and (\ref{HF chemical pot}), the semiclassical critical value of the interaction strength at finite temperature is 
\begin{equation}
    g_{2D,c}= \frac{2\pi \hbar^2}{m} \frac{1}{1-e^{-\beta\frac{\pi\hbar^2}{m}n}}.
    \label{semiclassical critical}
\end{equation}

The numerical solution of Eq. (\ref{Stoner criterion g critical}) is reported in Fig. \ref{fig 2} (see Appendix \ref{Appendix B} for details about the numerical calculations). In particular, in Fig. \ref{fig 2}-(a), the dimensionless critical interaction strength $g_{2D,c}$  is plotted as a function of the total number of fermions $N$. A peculiar peak structure appears at low temperatures, where pronounced and narrow peaks arise at magic numbers. Between shell closures, the critical interaction decreases as the temperature is reduced and approaches zero in the limit $T\to0$, since in a partially filled shell the degeneracy allows the gas to polarize without any kinetic-energy cost. Conversely, as $T \to 0$, for the magic numbers (completely filled shells), the critical interaction diverges: the energy gap separating neighboring angular momentum shells prevents polarization unless fermions are excited to the next available level, which is, however, forbidden at sufficiently low temperatures by the step-like Fermi-Dirac distribution.

This behavior can be understood by considering the $T \to 0$ limit of Eq. (\ref{Stoner criterion g critical}). In particular, for $\beta \to +\infty$ the function $\beta \operatorname{sech}^2(\beta x/2) \to 4 \delta(x)$. Hence, the $T=0$ interaction strength critical value is given by
\begin{equation}
 g_{2D, c} = \frac{4\pi R^2}{ \sum_{l=0}^{+\infty} (2l+1)\delta\left( \mathcal{E}_l-\mu^{\parallel}_\sigma+g_{2D}\frac{n}{2}  \right)} 
 \label{critical g T=0}.
\end{equation}
For a partially filled shell, the chemical potential coincides with the Hartree-Fock Fermi energy, Eq. (\ref{Fermi energy interacting unb}). Therefore, the delta function is satisfied precisely for the $l = l_F$ term in the summation. This yields a formally divergent denominator, which immediately forces the critical interaction to vanish, $g_{2D,c} \to 0$. On the other hand, for a completely filled shell (as discussed in the first Section), the chemical potential is not well defined, and in general it does not coincide with any discrete energy level. Instead, it is constrained within the energy gap between the Fermi shell and the lowest unoccupied shell. As a result, the argument of the delta function in Eq. (\ref{critical g T=0}) is always zero for any integer $l$, causing a divergence in the critical interaction strength, $g_{2D,c} \to +\infty$, as overcoming the energy gap would require a finite energy penalty.

As the temperature increases, the peak structure is gradually washed out, and the curves converge toward the semiclassical result, as the discreteness of the energy spectrum can be neglected and excited states become thermally available, similarly to what discussed in the non-interacting thermodynamics. 

The critical interaction strength at $T=0$ thus depends in a non-trivial way on the number of fermions on the sphere, unlike what happens in the flat two-dimensional gas, Eq. (\ref{semiclassical critical}), where at $T=0$ is density-independent and is given by $g_{2D,c}=2\pi\hbar^2/m$.

\begin{figure}{h}

    \centering
    \includegraphics[width=1.0\linewidth]{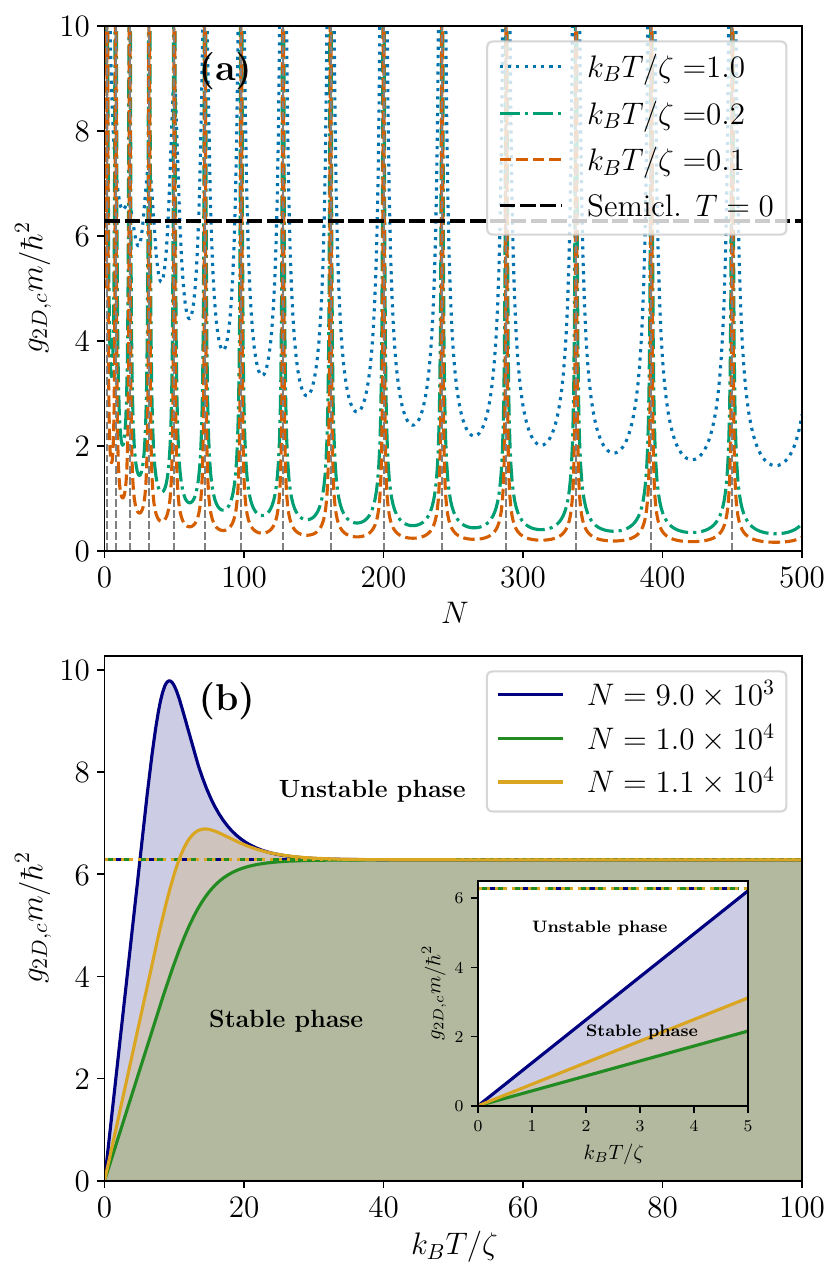}
    \caption{\small{ Repulsively interacting Fermi gas. The quantities are made dimensionless introducing the sphere energy scale $\zeta=\hbar^2/(mR^2)$, as explained in the main text.} (a) Dimensionless critical interaction strength $g_{2D,c} m/\hbar^2$ as a function of the number of fermions on the sphere $N$ for different temperatures: $k_BT/\zeta=1.0$ (blue  dotted line), $k_BT/\zeta=0.2$ ( green dash-dotted line) and $k_BT/\zeta=0.1$ ( orange dashed line).  The gray dashed vertical lines indicate the $N$ magic numbers, i.e the shells closures. (b) Dimensionless critical interaction strength $g_{2D,c} m/\hbar^2$ as a function of the dimensionless temperature $k_BT/\zeta$ for different number of fermions $N$, namely $N=9.0\times10^3$ (blue solid line), $N=1.0\times10^4$ (green solid line) and $N=1.1\times10^4$ (gold solid line). The colored plane portions correspond to the $g_{2D}$ values for which the spin-balanced solution is stable, namely for which Eq. (\ref{STONER CRITERION}) is not satisfied. The solid lines represent the critical coupling strength $g_{2D,c}$ above which Eq. (\ref{STONER CRITERION}) is satisfied. The semiclassical curves (provided by Eq. (\ref{semiclassical critical}), dashed lines) are also plotted as a reference. In the inset we highlight the convergence to zero of $g_{2D,c}$ as $T\to 0$. }
    \label{fig 2}
\end{figure}
Fixing the particle number $N$ and solving the number equation (\ref{HF equilibrium densities}) numerically, simultaneously with Eq. (\ref{Stoner criterion g critical}), allows us to determine the temperature dependence of the critical interaction strength and to plot the dimensionless gas instability phase diagram, which is shown in Fig. \ref{fig 2}-(b). At high temperatures, the behavior is indistinguishable from the semiclassical solutions, while shell effects manifest themselves at sufficiently low temperatures (when the discreteness of the spectrum cannot be ignored). Moreover, the numerical results confirm that $g_{2D,c} \to 0$ as $T \to 0$, except for magic numbers where $g_{2D,c} \to +\infty$. Finally, in the phase diagram, stability (instability) regions appear above (below) the semiclassical solution due to shell effects already observed in the non-interacting case, which strongly characterize the gas behavior at low temperatures. 

 As mentioned previously, the problem of a repulsive Fermi gas on a spherical surface at zero temperature has already been addressed in the literature. It is therefore appropriate to compare the results obtained here with the findings of Ref. \cite{phasesepsphere} (notice how $g_{2D}$ corresponds exactly to the dimensional parameter $\tilde{g}_{12}$ of Ref. [28]. Consequently, the relationship with their dimensionless parameter $g_{12}$ is simply $g_{2D} = g_{12} {2\pi \hbar^2}/{m}$). In that work, the authors determine the global energy minimum by allowing the spatial density of a completely filled shell configuration (specifically, $N=200$) to vary across the sphere. In contrast, our criterion strictly evaluates the local stability of the homogeneous phase against uniform density fluctuations for an arbitrary shell configuration. We demonstrate that for magic numbers, the homogeneous spin-balanced state (the “uniform" phase in \cite{phasesepsphere}) remains stable (i.e. a minimum) of the grand potential against uniform fluctuations for any finite $g_{2D}$. However, as shown in Ref. \cite{phasesepsphere}, for a sufficiently large interaction strength, the true global ground state of the system transitions to a phase-separated configuration (the  ``two-chunk” phase). However, our findings do not contradict those of Ref. \cite{phasesepsphere}; instead, they offer a complementary perspective (at least in the specific configurations in common between our studies). While they characterize the global ground state boundary, we determine the threshold for local instability, deducing from the comparison of results that the uniform state becomes metastable (i.e. a local minimum) in the strongly interacting regime where phase separation dominates. Future studies could bridge these approaches by investigating local, non-uniform density fluctuations on the sphere at finite temperatures across arbitrary shell occupancies.

\section{Conclusions}
\label{Conclusions}
In this paper, we have studied the properties of a fermion gas confined on a spherical surface, highlighting how the combination of intrinsic features of the system - Fermi statistics and finite size due to its constant curvature - leads to peculiar quantum effects and affects the low-temperature properties of the gas compared to the standard flat two-dimensional case already in the simple non-interacting case. For the more challenging case of the repulsive interacting gas, we have derived the Stoner instability criterion within a mean-field HF approximation, using an effective functional integration formalism to tackle the problem at finite temperature. The results obtained corroborate what was found in the non-interacting case, that is, how the non-trivial features of the spherical surface, on which the gas is confined, influence its behavior. We stress that the curvature-induced effects discussed are finite-size and finite-particle-number effects, as they naturally vanish when approaching the thermodynamic limit (where the flat two-dimensional behavior is recovered).

Our theoretical results can be experimentally tested using ultracold fermionic atoms confined in spherical bubble traps in microgravity conditions (for instance, in the NASA CAL laboratory) to avoid the accumulation of atoms on the lower side of the trap due to gravity. The two spin species components could be realized through two hyperfine states of fermionic ultracold atoms, e.g. using $^{6}\mathrm{Li}$ or $^{40}\mathrm{K}$ atoms. 
However, the experimental detection of the peculiar shell effect will definitely pose a significant experimental challenge. Indeed, considering a sphere of radius 
$R \approx 10 \ \mu \text{m}$, with $N \approx 10^4$ atoms, the temperatures required to observe the shell effect highlighted in Figs. \ref{fig 1}-\ref{fig 2} are on the order of $T \approx 1\ \text{nK}$, which is at the edge of the current experimental capabilities. One way to make these effects observable at higher and more accessible temperatures would be to reduce the radius of the sphere. With $R \approx 1 \ \mu \text{m}$, $T \approx 100\ \text{nK}$, which can be achieved in the laboratory \cite{Carollo2022}. 
Nevertheless, while current experiments do not allow the direct observation of shell effects, they would still enable the test of the semiclassical predictions. In light of the rapid progress and growing interest in this field, we expect that these experimental limitations will be overcome in the coming years as technologies and trap designs continue to advance. 

From a theoretical perspective, a natural extension of this work (in addition to what has already been highlighted) would be the study of attractive Fermi gases on the surface of a sphere. In particular, a BCS-like system could be studied, and investigation of the BCS-BEC crossover could provide insight into how the interplay between curvature and interactions affects the physical properties of the gas, as the system evolves from a weakly paired (BCS) regime to a strongly paired (BEC) regime \cite{Randeria81}. This study could also include an investigation of superfluid features of the gas, such as vortex dynamics, and of the Berezinskii-Kosterlitz-Thouless transition \cite{JMKosterlitz_1973, Berezinskii}. Furthermore, along the lines of what is discussed in this paper, attractively interacting spin-imbalanced mixtures could be considered to explore whether exotic pairing mechanisms, such as the Fulde-Ferrell-Larkin-Ovchinnikov state \cite{Sheevy15} or a 2D Fermi polaron \cite{polaron1, Polaron2}, might emerge on the sphere.

\section*{Acknowledgments} 
The authors are grateful to Cesare Vianello, Andrea Tononi and Giacomo Gradenigo for helpful discussions and valuable suggestions.
LF acknowledges the project “Frontiere Quantistiche” within the 2023 funding program ‘Dipartimenti di Eccellenza’ of the Italian Ministry for Universities and Research. AB acknowledges the European Quantum Flagship Project PASQuanS 2.
LS is partially supported by the European Union-NextGenerationEU within the National Center for HPC, Big Data, and Quantum Computing [Project No. CN00000013, CN1 Spoke 10: Quantum Computing]. 
LF, AB, and LS 
are partially supported by
Iniziativa Specifica “Quantum" of Istituto Nazionale di Fisica Nucleare (INFN) and the PRIN 2022 project “Quantum Atomic Mixtures: Droplets, Topological Structures, and Vortices”.

\appendix

\section{Justification of the grand canonical approach}
\label{Appendix D}

In principle, a gas of fermions constrained on the surface of the sphere is a closed system with a conserved number of particles, which should ideally be described by either a micro-canonical or canonical ensemble. However, operating directly in the micro-canonical (or canonical) ensemble within a quantum statistical mechanics approach is computationally and analytically prohibitive, as the known constraints $\sum_\alpha n_\alpha = N$ appear when calculating the partition function, where $\alpha$ indicates the collection of quantum numbers. This is a known problem and, for this reason, the standard approach at finite temperature is to use the grand canonical ensemble, which allows us to calculate the partition function. In our case, we can evaluate the error introduced by operating in the grand canonical ensemble by estimating the fluctuations of the number of particles in this ensemble, as the variance $(\Delta N)^2$ of the number of particles. This is given by the standard statistical mechanics relations 
\begin{equation}
    (\Delta N)^2 = \langle N^2 \rangle - \langle N \rangle^2 = k_B T \left( \frac{\partial \langle N \rangle}{\partial \mu_\parallel} \right)_{T,V},
    \end{equation}
where $\braket{N}$ is the grand canonical average. Ideally, this corresponds to fixing $N=\braket{N}$ in the canonical ensemble, where the number of particles does not fluctuate. The relative error introduced by this assumption is estimated by ${\Delta N}/{\langle N \rangle}$, and is expected to vanish in the thermodynamic limit. In the non-interacting Fermi gas, the total (average) number of particles calculated within the grand canonical ensemble is given by Eq. (\ref{Average number}). Differentiating explicitly with respect to the chemical potential leads to
\begin{equation}
    (\Delta N)^2 = 2\sum_{l=0}^{+\infty} (2l+1) f(\mathcal{E}_l) \big[1 - f(\mathcal{E}_l)\big],
\end{equation}
where $f(\mathcal{E}_l)$ is the Fermi-Dirac distribution. We can analyze two regimes. First, if $\mathcal{E}_l \ll \mu_\parallel$ (deep $l$ shells, occupied at very low temperatures), then $f(\mathcal{E}_l) \simeq 1$, which gives a nearly zero contribution to the sum. Second, when $\mathcal{E}_l \gg \mu_\parallel$ (empty $l$ shells above the Fermi energy), $f(\mathcal{E}_l) \simeq 0$, these shells also do not contribute. Therefore, only shells close to the Fermi level, within a window of $2k_B T$, significantly contribute to the summation. Let us assume that $\mu_\parallel$ coincides exactly with the energy of the Fermi shell $l_F$ (which is a valid assumption in the temperature regime in which we are interested, regardless of the occupation of the shell) and that the only shell that contributes to the summation (as a first approximation) is $l_F$. The variance becomes then
\begin{equation}
(\Delta N)^2 \simeq 2(2l_F + 1) f(\varepsilon_F) \big[1 - f(\varepsilon_F))\big] \propto l_F    
\end{equation}
(for instance, if the Fermi level is half populated then $f(\varepsilon_F)=1/2$, which corresponds to the maximum possible variance, hence the ``worst case scenario")
The total number of particles, instead, scales approximately as the square of the Fermi level
\begin{equation}
\braket{N} = 2\sum_{l=0}^{l_F} (2l+1) = 2(l_F+1)^2 \simeq 2l_F^2.
\end{equation}
Hence $l_F \propto \sqrt{\braket{N}/2}$, so that
 $ (\Delta N) \propto  \sqrt{l_F} \propto ({N/2})^{1/4}$.
Finally, the relative error scales as (as a first approximation)
\begin{equation}
    \frac{\Delta N}{\braket{N}} \propto \frac{1}{\braket{N}^{3/4}}.
\end{equation}
The fluctuations of the number of particles for this system in the grand canonical ensemble are strongly suppressed for the non-interacting gas, so that even for $N=100$ the relative error we commit is of the order of $3 \%$. Hence, the use of the grand canonical ensemble is justified, as in practice we are interested in $N \simeq 10^4$, which is close to the experimentally achieved number of particles.\\

Furthermore, turning on the repulsive interaction would result in an additional suppression of these ``fictitious" (in the sense that shouldn't be physically possible in the system under study) fluctuations because, while in the non-interacting gas the particles are - in principle - free to float  between the system and the fictitious particle reservoir, with a repulsive interaction between the atoms the system would oppose the addition of new particles, since it would cost interaction energy to push them against those already present.

\section{Details on the numerical calculations}
\label{Appendix B}

To numerically solve the dimensionless Eq. (\ref{Average number}) and explicitly find $\mu_\parallel(N)$ the procedure is straightforward: given a set of parameters $\{\beta$, $N_{target}\}$, which correspond to a point in the curves plotted in Fig. \ref{fig 1}, we simply numerically inverted equation (\ref{Average number}), solving numerically the equation $N(\mu_\parallel)-N_{target}=0$ to determine the chemical potential $\mu_\parallel$ corresponding to the target total particle number $N_{target}$ (with $\beta$ fixed during this process). The series $N(\mu_\parallel)$ (right-hand side of Eq. (\ref{Average number}) ) was evaluated adopting a cutoff $l_{cut}$, $\sum_{l=0}^{+\infty} \to \sum_{l=0}^{l_{cut}} $ large enough such that for every temperature considered in the calculations, we did not cut from the summation thermally available states. We found some difficulties when working at fixed $N$ in the case in which it corresponded to a fully occupied Fermi shell. In this case, the function $N(\mu_\parallel)$ at low temperature ($T\ll1)$ is flat for $N=N_{target}$ (see Fig. \ref{fig 2}), because of the energy gap in the single particle spectrum, so that the algorithm used to solve the equation numerically performs very poorly.

To study the stability of the gas and the transition from an unpolarized to a polarized gas, Eqs. (\ref{STONER CRITERION}) and (\ref{HF equilibrium densities}) have been solved numerically. They describe respectively the stability condition and the number equation which fix the total number of fermions on the sphere. Since an expression of the form $e^{\beta \varepsilon_l(n)}$ appears in the exponential of both equations, we implemented an alternative algorithm to an otherwise computationally expensive self-consistent calculation. Defining 
\begin{equation}
    \mu_{eff}=\mu_\parallel-g_{2D} \frac{n}{2}
\end{equation}
which appears only in the right-hand side of Eqs. (\ref{STONER CRITERION}) and (\ref{HF equilibrium densities}), the system can be reduced into the form 
\begin{equation}
    \begin{cases}
  g_{2D, c} = \frac{1}{\beta}\frac{16\pi R^2}{ \sum_{l=0}^{l_{cut}} (2l+1)\ \operatorname{sech}^2\big(\beta (\mathcal{E}_l-\mu_{eff}) \big)} \\
  \frac{N}{2} = \sum_{l=0}^{l_{cut}} \frac{(2l+1)}{e^{\beta (\mathcal{E}_l-\mu_{eff})}+1}
  \label{system of stoner}
\end{cases}.
\end{equation}
where $n=N/(4\pi R^2)$. In this way, the dependence on the exponential of the interaction term is canceled, avoiding a self-consistent solution and allowing us to solve the equations separately as a function of $\mu_{eff}$.
The real chemical potential can then be found by using the $\mu_{eff}$ definition.
This formulation allows us to construct parametric plots $N$ versus $g_{2D,c}$ using a common grid of $\mu_{eff}$ values to solve Eq. (\ref{system of stoner}).
By choosing to operate at fixed particle number $N_{target}$, instead, we can invert the number equation to find the equilibrium $\mu_{eff}$ value and, substituting it inside the other equation of (\ref{system of stoner}), obtain the critical interaction strength without the need of a time-consuming self-consistent calculation. We stress that we have verified that performing a self-consistent calculation leads to the same numerical results.  During the calculations, we have used a  value of $l_{cut}$ varying between 500 and 1000, depending on the average number of atoms considered. For each calculation, we ensured that this cutoff was sufficiently large to accurately describe the system, so that changing it did not alter the estimated values.

\section{Dimensional reduction}
\label{Appendix A} 

We propose a dimensional reduction procedure inspired by the approaches discussed in Refs. \cite{Salasnich2002, Moller2020}. We explicitly consider a three-dimensional shifted harmonic potential depending only on the radial coordinate,
\begin{equation}
    U(r)=\frac{1}{2}m\omega_\perp^2(r-R)^2
    \label{ harm conf pot}.
\end{equation}
This choice is physically motivated as, in experiments with low dimensional configurations, the confinement induced by optical-magnetic potentials is not perfect and the system always presents a finite thickness rather than being an ideal surface. The same happens for bubble trap setups where the atoms form a thin shell around the ideal spherical surface. The characteristic confinement length $R$ around the potential minimum is
\begin{equation}
    \ell_\perp=\sqrt{\frac{\hbar}{m \omega_\perp}}
    \label{ oscill lenght}.
\end{equation}
 As $\ell_\perp/R \to 0$ the harmonic radial confinement thickness shrinks, concentrating the particles near $r = R$. This will be the limit we consider to recover from a three-dimensional description an effective action on the sphere. We start by rewriting the three-dimensional Grassmann field, separating the radial and the angular part as
\begin{equation}
    \Phi_\sigma(r, \theta, \varphi, \tau)=\chi(r) \psi_\sigma(\theta, \varphi, \tau).
    \label{field decomposition}
\end{equation}
As proposed in \cite{Salasnich2002}, the radial field can be written as
\begin{equation}
    \chi(r)= \mathcal{N} e^{-\frac{(r-R)^2}{2\ell_\perp^2}}
    \label{ansatz}
\end{equation}
where $\mathcal{N}$ is a normalization factor to be determined. In full generality, the radial field may result from a superposition of different radial wavefunctions from the single particle problem. We assume that the particles can occupy only the radial ground state of the single particle problem, i.e. $\hbar \omega_\perp \gg \mu, k_BT, \mathcal{E}_l$, so that radial excitations are frozen out. The normalization factor $\mathcal{N}$ is calculated imposing 
\begin{equation}
   \int_0^{+\infty} dr\ r^2 |\chi(r)|^2=1.
   \label{norm cond}
\end{equation}
After some tedious calculations, $\mathcal{N}$ is determined as
\begin{equation}
    \mathcal{N}=\frac{1}{\sqrt{\frac{1}{2} \ell_\perp^2 R e^{-\frac{R^2}{\ell_\perp^2}} + \frac{\sqrt{\pi}}{4} \ell_\perp (2R^2 + \ell_\perp^2) \left(1 + \operatorname{erf}\left( \frac{R}{\ell_\perp} \right) \right)}}
    \label{Normalization factor}
\end{equation}
where $ \operatorname{erf}(x) = \frac{2}{\sqrt{\pi}} \int_0^x e^{-t^2} dt$. Notice that, in the limiting case in which $\ell_\perp/R \to 0$, the normalization factor can be approximated at first order in $\ell_\perp$ as
\begin{equation}
    \mathcal{N}\simeq \frac{1}{(\pi)^{\frac{1}{4}}R \sqrt{\ell_\perp}},
\end{equation}
and since
\begin{equation}
    \frac{1}{\sqrt{\pi} \ell_\perp}e^{-\frac{(r-R)^2}{\ell_\perp^2}} \overset{\text{$ \ell_\perp/R \to 0$}}{\longrightarrow} \delta(r-R), 
\end{equation}
in such limit the radial wavefunction reduces to a Dirac delta centered at $r=R$, which means that all the particles now live on the surface of a sphere. We are now ready to integrate out the radial degrees of freedom and obtain an effective path integral on the sphere.

Let us focus first on the non-interacting three-dimensional Euclidean Lagrangian density
\begin{align}
    \mathscr{L}_0[\bar{\Phi}_\sigma,{\Phi}_{\sigma} ] =&\sum_{\sigma = \{ \uparrow, \downarrow \}} 
\bar{\Phi}_\sigma 
\left(  \hbar \frac{\partial }{\partial \tau} - \frac{\hbar^2}{2m}\nabla^2 +U- \mu\right)
{\Phi}_{\sigma} 
\label{three dim lagrangian}
\end{align}
where the field $(r,\theta, \varphi, \tau)$ dependence is understood. Combining Eqs. (\ref{field decomposition}), (\ref{ansatz}) and (\ref{Normalization factor}), and writing the Laplacian in spherical coordinates
\begin{equation}
    \nabla^2 = \frac{1}{r^2}\frac{\partial }{\partial r} \left( r^2 \frac{\partial}{\partial r} \right) - \frac{\hat{L}^2}{\hbar^2 r^2} 
    \label{Nabla spherical},
\end{equation}
 we evaluate the action of the operators inside the square bracket of Eq. (\ref{three dim lagrangian}) on the fields. We obtain
\begin{multline}
    \chi^*(r)\bar{\psi}_\sigma(\theta, \varphi, \tau) 
\Big\{ \chi(r) \hbar \frac{\partial }{\partial \tau} \\  -\frac{\hbar^2}{2m\ell_\perp^4} \chi(r)\left( (r-R)^2-\ell_\perp^2(3-2\frac{R}{r})\right)  +\chi(r)\frac{\hat{L}^2}{2mr^2}\\+ \frac{\chi(r)}{2}\omega_\perp^2(r-R)^2 - \chi(r)\mu  \Big\}  {\psi}_{\sigma}(\theta, \varphi, \tau)
\label{full 3dim expression}.
\end{multline} 
Inserting it inside the three-dimensional Euclidean action
\begin{equation}
    \mathcal{S}_0[\bar{\Phi}_\sigma,{\Phi}_{\sigma} ]=\int_0^{\hbar \beta} d\tau \int d^3\vec{r} \ \mathscr{L}_0 [\bar{\Phi}_\sigma,{\Phi}_{\sigma} ]
\end{equation}
we integrate over the radial coordinate in the limit $\ell_\perp/R \to 0$, obtaining $\mathcal{L}_0$, the effective non-interacting Lagrangian on the surface of the sphere. Explicitly
\begin{equation}
    \int_0^{\infty}|\chi(r)|^2 dr \overset{ \ell_\perp/R \to 0}{\simeq} \frac{1}{R^2},
\end{equation}
and
 \begin{multline}
  \int_0^{\infty}r^2 dr|\chi(r)|^2 \Big\{-\frac{\hbar^2}{2m\ell_\perp^4} \Big( (r-R)^2 -\ell_\perp^2(3-2\frac{R}{r})\Big)\\ + \frac{1}{2}m\omega_\perp^2(r-R)^2\Big\}\overset{\text{$ \ell_\perp/R \to 0$}}{\simeq}\frac{\hbar^2}{2m\ell_\perp^2} =\frac{\hbar \omega_\perp}{2}  
 \end{multline}
which contributes only through a constant to the Euclidean action, that can be rewritten using Eq. (\ref{ oscill lenght}). It corresponds to the zero-point energy of the harmonic oscillator in the radial direction, consistent with what was obtained in Ref. \cite{Moller2020}. The other terms of Eq. (\ref{full 3dim expression}) do not depend on the radius, allowing us to exploit the normalization (\ref{norm cond}) of the radial field to integrate over $r$. This constant can be reabsorbed into the chemical potential as a constant shift, and we can think of it as the energy needed to add a particle in the thin spherical shell due to the presence of the harmonic trap. In this sense, it can be interpreted as a radial chemical potential contribution $\mu_\perp$, such that the chemical potential on the sphere $\mu_\parallel$ is given by
\begin{equation}
    \mu-\frac{\hbar^2}{2m\ell_\perp^2}=  \mu_{\parallel} .
\end{equation}

Finally, the effective Euclidean non-interacting action on the sphere  can be written as
\begin{equation}
    S_0[ \bar{\psi}_\sigma, {\psi}_\sigma ]=\int_{0}^{\hbar\beta} d\tau \int_0^\pi \sin(\theta) d\theta \int_0 ^{2\pi} d\varphi \ \mathcal{L}_0 [ \bar{\psi}_\sigma, {\psi}_\sigma ]
\end{equation}
where the non-interacting Euclidean Lagrangian density per solid angle $\mathcal{L}_0 [ \bar{\psi}_\sigma, {\psi}_\sigma ]$ is given by
\begin{equation}
    \mathcal{L}_0 [ \bar{\psi}_\sigma, {\psi}_\sigma ] =\sum_{\sigma = \{ \uparrow, \downarrow \}} 
\bar{\psi}_\sigma 
\left(  \hbar \frac{\partial }{\partial \tau} + \frac{\hat{L}^2}{2mR^2} - \mu_{\parallel}\right)
{\psi}_{\sigma}.
\label{ non int lagrangian density sphere}
\end{equation}
This is the natural generalization to fermions to the path-integral formalism used in Ref. \cite{BECsphere} to study a gas of bosons. 

Let us now turn to the interacting part of the three-dimensional Euclidean Lagrangian density, namely
\begin{equation}
    \mathscr{L}_I [ \bar{\Phi}_\sigma, {\Phi}_\sigma ] = \frac{1}{2} \sum_{ \sigma, \sigma' = \{\uparrow, \downarrow \}} 
\int d^3 r' \
\bar{\Phi}_\sigma
\bar{\Phi}_{\sigma'} 
V_{\sigma, \sigma'}
{\Phi}_{\sigma'}
{\Phi}_\sigma
\end{equation}
where $V_{\sigma, \sigma'}(\vec{r}, \vec{r}^\prime)$ is a three-dimensional contact interacting potential
\begin{equation}
    V_{\sigma, \sigma'}(\vec{r},\vec{r}\,^{\prime})=g_{3D}\ \delta^{(3)}(\vec{r}-\vec{r}\,^{\prime}) (1-\delta_{\sigma, \sigma'}).
\end{equation}
Here $ g_{3D}={4\pi\hbar^2 a_s}/{m}$ is the three-dimensional interaction strength, related to experiments through the s-wave scattering length $a_s$. The contact interacting potential on the sphere can be modeled as 
\begin{equation}
  V_{\sigma, \sigma'}(\theta, \varphi; \theta', \varphi')=g_{\parallel}\delta(\cos\theta-\cos\theta')\delta(\varphi-\varphi')(1-\delta_{\sigma,\sigma'})
\end{equation}
where $g_{\parallel}$ defines the interaction strength on the sphere surface. If $g_{\parallel}>0$, then the interaction between the atoms is repulsive; if instead $g_{\parallel}<0$, then the interaction between the atoms is attractive. The interaction strength $g_{\parallel}$ can be linked to the three-dimensional coupling constant $g_{3D}$ by integrating out the radial coordinate and taking the limit $\ell_\perp/R \to 0$. Thus, we define
\begin{equation}
 g_{\parallel}= g_{3D}\int_0^{+\infty} dr\ r^2 |\chi(r)|^4  \overset{\ell_\perp/R \to 0}{\simeq} \frac{g_{3D}}{\sqrt{2\pi}\ell_\perp R^2}. 
    \label{interaction reduction}   
\end{equation} 
so that the interacting Euclidean Lagrangian density on the sphere surface becomes
\begin{equation}
    \mathcal{L}_I= g_{\parallel}\ \bar{\psi}_\uparrow  \bar{\psi}_\downarrow   \psi_\downarrow  \psi_\uparrow 
    \label{int lagr dens}.
\end{equation}
and the full Lagrangian density is simply $ \mathcal{L}= \mathcal{L}_0+ \mathcal{L}_I$.
Again, this is the fermion generalization to the formalism for bosons already used in several papers (e.g \cite{BECsphere}). Moreover, from Eq. (\ref{interaction reduction}) we are able to recover the result of the dimensional reduction procedure of Ref. \cite{Moller2020}, defining the two-dimensional interaction strength 
\begin{equation}
    g_{2D}=g_\parallel R^2.
\end{equation}

\section{Regularization scheme of divergent Matsubara frequency}
\label{Appendix C}

The Matsubara frequency summation considered in this work is of the form  
\begin{equation}
\sum_{s=-\infty}^{+\infty} \ln\left(\beta(-i\omega_s\hbar+\mathcal{E}) \right),
\end{equation}
where $\omega_s={(2s+1)\pi}/{\hbar\beta}$, $s \in \mathbb{Z}$, are the fermionic Matsubara frequencies and $\mathcal{E}$ is an energy that does not depend on the index $s$. This sum is divergent but, as shown in Ref. \cite{AltlandSimons2010}, it can be regularized by introducing a convergence factor $e^{i\omega_s 0^+}$, which naturally arises from the implicit time-ordering of the Grassmann fields in the path integral.  
Following the prescription discussed in Ref. \cite{AltlandSimons2010}, a $e^{i\omega_s \delta}$ factor is included inside the summation and the limit $\delta \to 0^+$ is taken at the end of the calculation
\begin{equation}
\sum_{s=-\infty}^{+\infty} \ln\left(\beta(-i\omega_s\hbar+\mathcal{E}) \right)e^{i\omega_s\delta}.
\end{equation}

To evaluate the sum, we considered the analytic continuation of the Fermi-Dirac function
\begin{equation}
g(z)=\frac{1}{1+e^{\beta \hbar z}}.
\end{equation}
It has simple poles at $z=i\omega_s$ with residues $\operatorname{Res}_{z=i\omega_s} g(z)=-1/(\beta\hbar)$.  
This allows us to rewrite the Matsubara summation as a contour integral in the complex plane
\begin{equation}
\sum_{s=-\infty}^{+\infty}f(i\omega_s)= -\beta\hbar \oint_\mathcal{C} \frac{dz}{2\pi i} g(z)f(z),
\end{equation}
with $f(z)=\ln(\beta(-\hbar z+\mathcal{E})) e^{\delta z}$ and $\mathcal{C}$ a closed contour which enclose the imaginary axis. 
By deforming the latter to an infinitely large circle, in such a way as to avoid the branch cut of the logarithm along the positive real axis, one finally is able to obtain the finite contribution
\begin{equation}
\sum_{s=-\infty}^{+\infty} \ln\left(\beta(-i\omega_s\hbar+\mathcal{E}) \right)e^{i\omega_s\delta} = \ln(1+e^{-\beta\mathcal{E}}),
\end{equation}
which is the final regularized Matsubara sum.

\bibliography{bibliography}
\clearpage

\newpage




\end{document}